# Crowd Capital in Governance Contexts

John Prpić
Beedie School of Business - Simon Fraser University

&

Prashant Shukla
Beedie School of Business - Simon Fraser University
Rotman School of Management - University of Toronto

**Abstract**

To begin to understand the implications of the implementation of IT-mediated Crowds for Politics and Policy purposes, this research builds the first-known dataset of IT-mediated Crowd applications currently in use in the governance context. Using Crowd Capital theory and governance theory as frameworks to organize our data collection, we undertake an exploratory data analysis of some fundamental factors defining this emerging field. Specific factors outlined and discussed include the type of actors implementing IT-mediated Crowds in the governance context, the global geographic distribution of the applications, and the nature of the Crowd-derived resources being generated for governance purposes. The findings from our dataset of 209 on-going endeavours indicates that a wide-diversity of actors are engaging IT-mediated Crowds in the governance context, both jointly and severally, that these endeavours can be found to exist on all continents, and that said actors are generating Crowd-derived resources in at least ten distinct governance sectors. We discuss the ramifications of these and our other findings in comparison to the research literature on the private-sector use of IT-mediated Crowds, while highlighting some unique future research opportunities stemming from our work.

1. Introduction

Launched on India's Independence Day in 2010 by the non-profit organization Janaagraha, IPaidaBribe[1] has collected nearly 25,000 reports of bribery across 645 Indian cities. Elsewhere, half the globe away, the Philadelphia police force has instituted the SafeCams program to leverage the digital cameras of their citizens to investigate crime in their municipality. In Abu Dhabi, the government has launched Cityguard, a mobile application for residents of the Emirate allowing the public to report incidents and submit complaints directly to the government. Similarly, in the UK, a social enterprise known as FixMyStreet has launched, resulting in tens of thousands of local problems being addressed by municipalities across the UK. In Syria, two American women, using the open source Ushahidi platform and a consortium of corporate, foundation, and individual funding, launched Women under Siege,[2] therein documenting hundreds of cases of sexual violence against Syrian women during the ongoing civil war.

In Mali, the French foreign services have launched 'L'aide Francaise au Mali' to track the status of their foreign aid projects in the country. In the United States, the Federal government has launched Challenge.gov platform to attempt to solve the most pressing problems facing federal agencies (Brabham, 2013). In Finland, a Finnish parliament standing committee including the Prime Minister as a member, recommends that the parliament should process ideas for legislative change emanating from a non-profit web portal known as the Open Ministry (Aitamurto, 2012). In the United States, the U.S. Patent and Trademark Office, in conjunction with the NYU Law School and several major patent-holding companies launch Peer to Patent enlisting a Crowd of volunteers to search for prior art (Brabham, 2013). In Iceland, a constitutional council of 25 people uses a Facebook page to seek popular input on their successive drafts of proposed constitutional changes (Burgess & Keating, 2013; Landemore, 2014). In the United States, the US Army launches ArmyCoCreate asking their soldiers in the field for ideas to be implemented by their rapid equipping force.

In all these numerous cases, and the many others not mentioned thus far, we see that individuals and organizations are using IT to engage Crowds for the purpose of creating resources to be used in the governance context. By any measure, the collaborative, technology-intensive paradigm of innovation, production, idea-generation and problem solving (Benkler, Roberts, Faris, Solow-Niederman, & Etling, 2013; de Vreede, Briggs, & Massey, 2009) has arrived in the governance context too. Ranging from health care (Kim, Lieberman, & Dench, 2014), intellectual property and legislation, to foreign aid (Bott, Gigler, & Young, 2014), law enforcement (The Swedish Program for ICT in Developing Regions, 2013) and military, we are beginning to see functions and issues that have traditionally been solely within the purview of the government apparatus now enlisting the aid of IT-mediated Crowds.

---

[1] http://www.ipaidabribe.com/
[2] https://womenundersiegesyria.crowdmap.com/

Given the central role of policy and political governance for the operation of 21$^{st}$ century nations and economies, the nascent arrival of the use of IT-mediated Crowds in the governance context signals an important change in the function, role, and reach, of political and policy governance. Unlike the corporate use of IT-mediated Crowds, largely aimed at narrow profit pursuit purposes, the use of IT-mediated Crowds in governance raises novel concerns at the intersection of the legislative, judicial, and executive branches of government, at all levels of government operations, and in all rule-bound nations. Therefore, given the importance and potential complexity of the use Crowds for governance, the nascent and rapid emergence of such applications in the governance context, and the conspicuous dearth of research in the area, our work begins to sketch the contours of this salient new research area by pioneering the first research effort demarcating the field.

In the ensuing sections of this paper, we will achieve these research aims by first reviewing the literature on IT-mediated Crowds in section # 2, and the governance context in section # 3, therein introducing the lenses that guide our data collection in section # 4. In section # 5 we illustrate the findings of our exploratory analysis, introducing and outlining some universal factors common to all IT-mediated applications in the governance context. In section # 6, we discuss the ramifications of our findings focusing on both the observed and the potential implications of the use of IT-mediated Crowds in the governance context, before concluding by outlining some important and unique research opportunities stemming from our work.

## 2. IT-Mediated Crowds

The Theory of Crowd Capital (TCC) perspective (Prpic, Shukla, Kietzmann, & McCarthy, 2015; Prpic & Shukla, 2013, 2014) is an organizational-level model outlining how and why organizations are using IT to engage Crowds for resource purposes. The Crowd Capital perspective captures the essence and dynamics of numerous substantive research areas including: Prediction Markets, Wikis, Citizen Science, Crowdsourcing, Crowdfunding, and Open Innovation platforms, and formulates a generalized model of resource generation from IT-mediated Crowds. In Figure # 1 below we outline a systemic perspective of the constructs of the Theory of Crowd Capital: Dispersed Knowledge is the antecedent condition (a Crowd), which is engaged by an Organization's Crowd Capability (Content, IT Structure, and Internal Processes), to generate the Crowd Capital resource for an Organization.

Crowd Capital is an organizational-level resource (knowledge or financial resources for example) generated from IT-mediated Crowds. From the perspective of the organization, an IT-mediated Crowd can exist inside of an organization, exist external to the organization, or some combination of the latter and the former. Crowd Capital resource generation is always an IT-mediated phenomenon, and is actuated through an organization's Crowd Capability - an organizational-level capability encompassing the three dimensions of; the form of content sought from a Crowd, an IT structure, and internal organizational processes.

The content dimension of Crowd Capability defines the form of the content sought from a Crowd (e.g. knowledge, information, data, money); the IT structure component of Crowd Capability indicates the technological means employed by an organization to engage a Crowd; and the process dimension of Crowd Capability refers to the internal procedures that the organization will use to organize, filter, and integrate the incoming Crowd-derived contributions. Crucially, IT structure can be found to exist in either Episodic or Collaborative form, depending on the interface of the IT used to engage a Crowd.

In the ensuing subsections we'll discuss each of these features of Crowd Capital theory, construct by construct.

**Figure # 1 – The Theory of Crowd Capital**

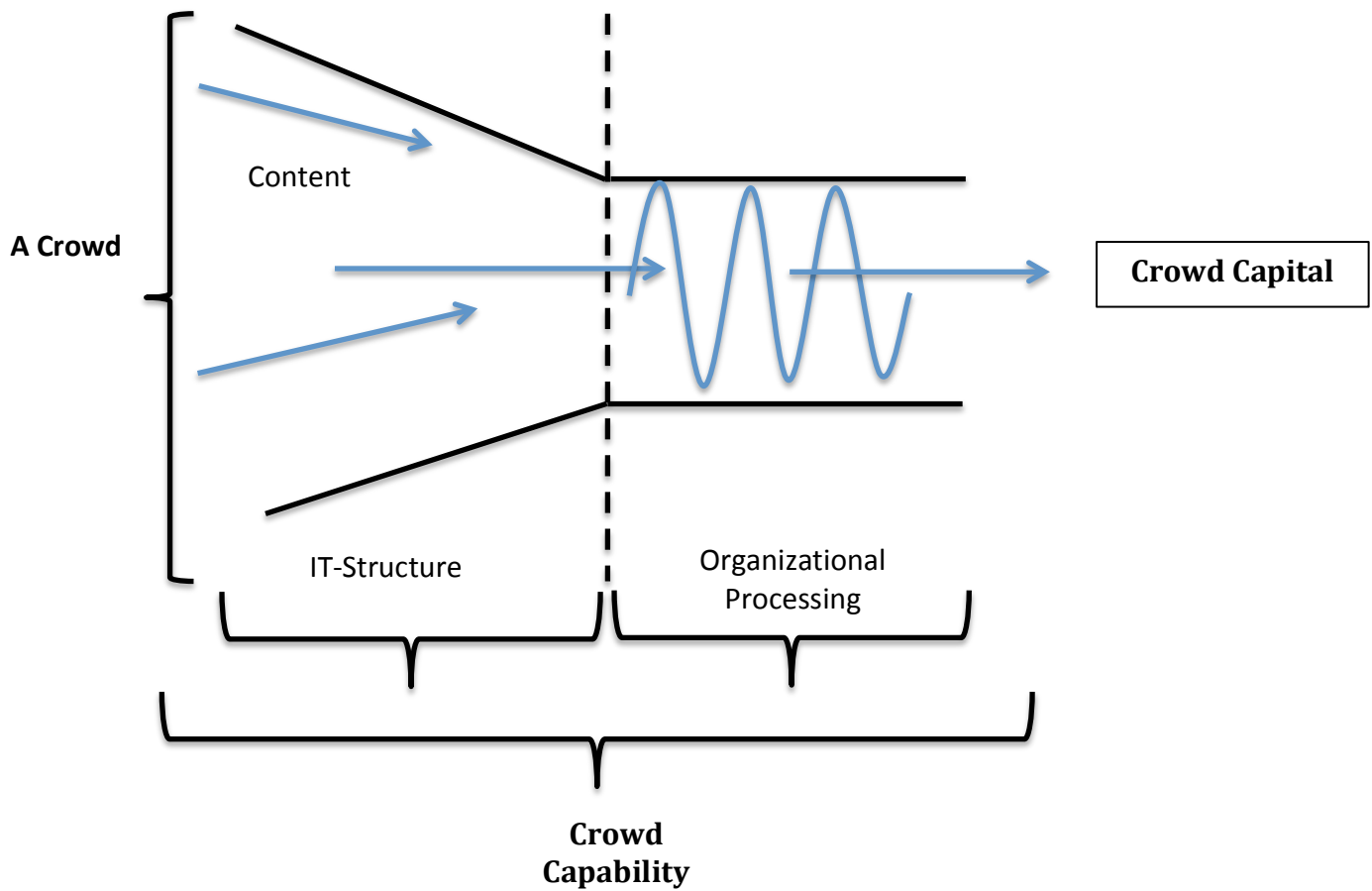

## 2.1 Dispersed Knowledge

Figure # 2 below, presents the major constructs of the TCC, with the dispersed knowledge as the antecedent construct of TCC. The existence of dispersed knowledge has been the subject of inquiry in economics for many years, and central to our understanding of dispersed knowledge is the contribution of F.A. Hayek, who in 1945 wrote a seminal work titled 'The Use of Knowledge in Society'.

In this work, for which Hayek was eventually awarded the Nobel prize, Hayek describes dispersed knowledge as "…the knowledge of the circumstances…never exists in concentrated or integrated form but solely as the dispersed bits of incomplete and frequently contradictory knowledge which all the separate individuals possess" (Hayek, 1945). In his conception: "…every individual…possesses unique information of which beneficial use might be made, but of which use can be made only if the decisions depending on it are left to him or are made with his active cooperation" (Hayek, 1945).

For Hayek, the existence of dispersed knowledge is the state of nature in society, "The problem which we meet here is by no means peculiar to economics but arises in connection with nearly all truly social phenomena… and constitutes really the central theoretical problem of all social science" (Hayek, 1945).

**Figure # 2 – The Theory of Crowd Capital—Constructs[3]**

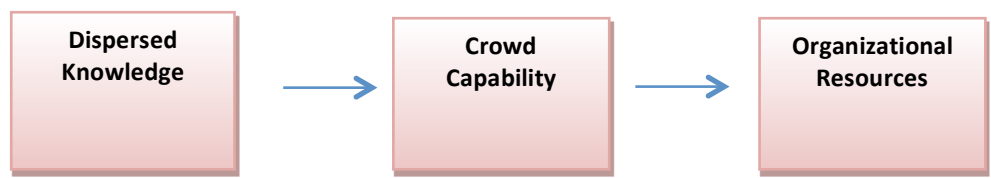

Therefore, in sum, dispersed knowledge in TCC describes why Crowds are useful for organizations to engage. A Crowd, comprised of collection(s) of independently-deciding groups or individuals (Reiter & Rubin, 1998; Surowiecki, 2005), represents a subset of all of the dispersed knowledge available in society writ large. And because dispersed knowledge changes moment to moment due to temporal factors, no Crowd, let alone any particular group or individual knowledge is ever static. Thus, every Crowd, even those comprised of the exact same individuals and groups, is always, and everywhere, unique from moment to moment. For the purposes of this particular study, we employ the dispersed knowledge construct to assist in our data collection and organization by focusing on the geographic dispersion of governance Crowds, grouped at a continental level.

---
[3] Adapted from Prpić and Shukla (2013; 2014) and Prpić et al. (2015)

## 2.2 Crowd Capability

Crowd Capability is an organizational-level capability that encompasses the structure, content, and process of an organization's engagement with a Crowd. The content dimension represents the form of content sought from a Crowd. Well-known forms of content that are currently being sought-out from Crowds include micro-tasks (Kulkarni, Can, & Hartmann, 2012), ideas and creativity (Brabham, 2013), money (Belleflamme, Lambert, & Schwienbacher, 2013) and technical innovative solutions (Lakhani & Panetta, 2007). The process dimension of Crowd Capability refers to the internal procedures that the organization will use to organize, filter, and integrate the incoming Crowd-derived content contributions. The IT structure component of Crowd Capability indicates the technological means employed by an organization to engage a Crowd, and crucially, IT structure can be found to exist in either Episodic or Collaborative form, depending on the interface of the IT used to engage a Crowd.

With Episodic IT structures, the members of the Crowd never interact with each other individually through the IT. A prime example of this type of IT structure is Google's reCAPTCHA (von Ahn, Maurer, McMillen, Abraham, & Blum, 2008), where Google accumulates significant knowledge resources from a Crowd of millions, though it does so, without any need for the Crowd members to interact directly with one another through the IT.

On the other hand, Collaborative IT structures require that Crowd members interact with one another through the IT, for resources to be generated. Therefore, in Collaborative IT structures, social capital must exist (or be created) through the IT for resources to be generated. A prime example of this type of IT structure is Wikipedia, where the Crowd members build directly upon each other's contributions through time.

This crucial distinction of IT structures, in turn, necessarily impacts the actual form of the interface of the IT artifact itself, and as such, we will employ it in the data collection and analysis to follow.

## 2.3 Crowd Capital

Crowd Capital is a heterogeneous organizational-level resource generated from IT-mediated Crowds. We label this newly emergent organizational resource as Crowd Capital because it is derived from dispersed knowledge (A Crowd), and because it is a key resource (a form of capital) for an organization, that can facilitate productive and economic activity (Nahapiet & Ghoshal, 1998). Like the other forms of capital (social capital, financial capital etc.), Crowd Capital requires investment (for example in Crowd Capability), and potentially leads to literal or figurative dividends, and therefore it is endowed with typical "capital-like" qualities. Further, in respect to TCC, the Crowd Capital construct is the outcome (or a potential outcome) of engaging IT-mediated Crowds.

For the purposes of this particular study, we employ the Crowd Capital construct to categorize the different types of resources being generated by actors in their use of IT-mediated Crowds in the governance context.

### 3. Governance & Governance Context

Governance theory as a definable body of political science research began by being concerned with the steering actions of political authorities as they deliberately attempt to shape socio-economic structures and processes (Mayntz, 1998), and has shifted to signify a change in the meaning of government, focusing on new processes by which societies are governed (Chhotray & Stoker, 2008; Rhodes, 1996; Stoker, 1998). The term governance, long equated with 'governing', the process aspect of government, thus complemented the institutional perspective of political studies. Recently, however, the term "governance" has been used in two other ways, both distinct from political guidance or steering (see Table # 1). It is important to distinguish these different and emergent meanings as changes in semantics may reflect changes in perception, and perhaps reflect changes in reality too (Mayntz, 1998).

It is now relatively common for the term governance to be used to indicate a new mode of governing that is distinct from the originating hierarchical control model. This change indicates a more cooperative mode of governing operations, where state and non-state actors participate in mixed public/private networks to direct society (Kooiman, 1993, 2003; Mayntz, 1998). Governance studies in the network approach, and as an alternative to hierarchical control, has been studied at the national and sub-national levels of European policy-making for example (Kohler-Koch & Rittberger, 2006), and prominently in international relations too (Dingwerth & Pattberg, 2006; Scholte, 2002).

The third evolution of the term governance is much more general in scope, due in part to its creation in Institutional economics. In this sense intended by this body of originating work, governance intimates the different forms of coordinating individual actions, and thus basic forms of social order (Mayntz, 1998). These ideas grew of transaction cost economics (Coase, 1937; Williamson, 1979), and it's analysis of market and hierarchies as alternative forms of economic organization. Williamson's typology in particular, was quickly extended to include other forms of social order, such as clans, associations, and networks (Hollingsworth & Lindberg, 1985; Powell, 1990). The net result of these works was that 'new' forms of coordination, different from both hierarchy and markets, led to the generalization of the term "governance" to cover all forms of social coordination - not only in the economy, but also in other sectors (Mayntz, 1998).

**Table # 1 – The Stages of the Evolution of the Theory of Political Governance (adapted from Mayntz, 1998)**

| Stage | Time of Appearance | Basic Idea |
|---|---|---|
| (1) | Early 1970s | Prescriptive theories of planning. |
| (2) | Later 1970s | Empirical studies of policy development (agenda setting, instrument choice, role of law, organizational context). |
| (3) | 1980's | Policy implementation. |

For the purposes of this work, we use of the notion of 'governance context' put forward recently by (Howlett & Lindquist, 2007), as a conceptual tool to organize our data collection and analysis context. In their view, the governance context:

> "…presumes that very different patterns or styles, and 'movements', of policy analysis can exist in different jurisdictions, policy sectors, and organizational contexts. These styles can include a penchant for the use of traditional 'generic' tools such as cost-benefit analysis, but can also, legitimately, include propensities for the use of alternate or complementary analytical techniques such as consultation and public or stakeholder participation, or long-standing preferences for the use of specific types of 'substantive' policy instruments or governance arrangements, such as regulation or public enterprises or the use of advisory commissions or judicial review…" (Howlett & Lundquist 2007).

We feel that the framing of the governance context concept used by Howlett & Lundquist (2007) captures all the elements of the three streams of governance theory outlined in Table # 1 (hierarchy, networks, empirical policy creation), while having the added benefit of capturing the more modern notion of tools (analytical or IT-based tools), and public participation that are key to our analysis.

Having now established the literature base for our data collection in the preceding sections, in the ensuing section we describe the details of our data collection process.

## 4. Data Collection

Through the use of secondary archival sources such as web pages, search engines, web alerts, mailing lists, social media, blogs, the general press, and the research literature, we assemble the only database that we are aware of, detailing endeavors where IT-mediated Crowds are being engaged solely in the governance context.

Our search and collection of the data began in September 2013, and continues as new applications emerge, and existing applications become known to us. As of this writing, our database includes 209 different applications. Once we become aware of an application, we investigate the source, generally a web page, to determine whether the application engages IT-mediated Crowds in a governance context, and if so we add it to our database, and categorizing the traits of the application along the dimensions of our Governance and Crowd Capital lenses.

For us, a governance context includes situations where IT-mediated Crowds are being implemented at any level of a sovereign government (federal, state, municipal) nationally or internationally. Non-state actors, such as individuals, non-profits, and private initiatives are also included in our dataset only if they aim at areas traditionally within the purview of the state apparatus.

For example, we include the use of IT-mediated Crowds by individual politicians if these uses are aimed at more than winning votes/elections. So while a politician using a Facebook or Twitter page to marshal his or her supporters would not be a part of our dataset (Hemphill, Otterbacher, & Shapiro, 2013), a member of a legislature using a wiki page or Reddit to solicit ideas relevant to legislation from constituents (or the public at large) would be included in our dataset (Mainka, Hartmann, Stock, & Peters, 2014) if it is an ongoing concern.

Similarly, smart city endeavours that draw on IT-mediated Crowds are included in our dataset (Nash, 2010; Seltzer & Mahmoudi, 2013), while E-Government initiatives (where some level of government allows its services to only be accessed online) are not (e.g. Criado, Sandoval-Almazan, & Gil-Garcia, 2013). Citizen Science initiatives are also excluded from our dataset, since we feel that resources generated from such scientific research is not directly in the governance context. Further, Microlending, Crowdfunding, and Crowd Journalism are similarly excluded from our dataset. In sum, we exclude all applications of IT-mediated Crowds targeted at business or business functions, and include only those applications targeted at generating resources within the purview of governance networks or the governing apparatus.

Along similar lines, it's important to note that Crowd Capital cannot exist with a "one-way" push of resources or information, whether IT-mediated or not. Developing or curating a web-based community, centred on the one-way communication of newsletters/updates/blog posts/mailing lists/web pages/blog comments etc., does not constitute generating Crowd Capital. In such cases, though these applications can be considered as IT-mediated communities, there is a more or less passive receipt of relatively pre-determined information resources, and few if any novel resources are created in the process. For the same reason that we exclude Crowd journalism applications from our dataset, we exclude these forms of IT-mediated communities, as in essence they represent forms of media content, which though important and useful in society are essentially an exercise of private/individual opinion, which we consider to be outside of the direct governance context.

Moreover, Crowd Capital generation is always, and only, an IT-mediated phenomenon, with only IT-mediated outcomes resulting, and thus many such web-based communities serve primarily to organize offline community involvement, meetings, hackathons, protests, social groups, community advocacy etc. Though these are endeavours are effective in generating offline 'collaborative governance' (Ansell & Gash, 2008; Newman, Barnes, Sullivan, & Knops, 2004), such collaborations are not novel, and are not solely IT-mediated in process or outcome, and thus we exclude the many such communities form our dataset, and thus our consideration too. This is not to say that these types of communities are not valuable, rather they are relatively mundane, and do not illustrate the unique, sometimes massively scaled, fast and dynamic resource generating capacities found in the forms of Crowd Capital creation, such as the forms or Crowdsourcing, Citizen Science, Crisis-mapping, Social media applications, and Wikipedia etc. All of which are new, and only IT-mediated.

Altogether, it must be noted that our dataset is most certainly not comprehensive, and we expect that it will continue to grow in size and shape as we continue to monitor the environment for the emergence of new instantiations, and to learn of extant one's that have thus far escaped our attention. As we detail above, we have attempted to be very vigilant in our exclusion of applications that do not meet our "pure play" strictures for both generating Crowd Capital, and doing so, solely within a governance context. Our efforts are an attempt to provide organization and clarity in this new and important domain, and we hope that our work is beneficial to practitioners and scholars alike in this respect.

In the ensuing section, we detail the findings of the exploratory data analysis undertaken with the assembled dataset described above.

## 5. Findings

As a fundamental starting point in our analysis of this new domain, we undertake some simple exploratory analysis of our assembled dataset, by calculating the relative distributions of the different IT-mediated applications for governance detailed in our dataset. The relative distributions are calculated within the categories delineated by our use of the Crowd Capital and Governance context lenses, used to organize our data collection. We discuss the categories in turn in each subsection below.

**5.1 Actors in the Governance Context**

As mentioned earlier, the governance context includes networks of actors involved in the governing of society, and thus we find it useful to begin to unpack this network of actors currently participating in the application of IT-mediated Crowd for governance.

We find a range of actors participating in the governance context, along the spectrum from private to public actors (Knill & Lehmkuhl, 2002; Mason, Kirkbride, & Bryde, 2007; Osborne, 2002). In Figure # 3, we present a graphical depiction of a spectrum of the different actors employing IT-mediated Crowds in the governance context, ranging from fully private actors on the left, to fully public actors on the right.

**Figure # 3 – Types of Actors Employing IT-mediated Crowds in the Governance Context**

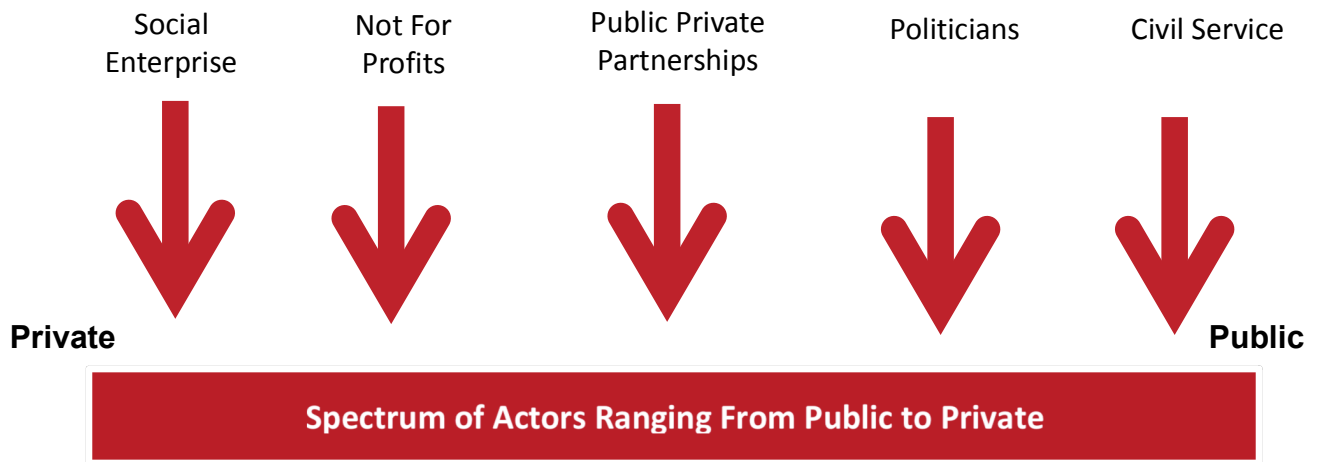

Of the 209 applications in our dataset, we find that social enterprise and non-profits have the highest percentages of occurrence in respect to the type of actor. Table # 1 below summarizes this information for the different types of actors implementing IT-mediated Crowds in the governance context.

**Table # 1: Percentages of each type of actors**

| Actor-Type | Percentage of Overall Dataset |
|---|---|
| Social Enterprise | 38% |
| Non-Profits | 29% |
| Civil Service | 16% |
| Public-Private Partnerships | 15% |
| Politicians | 02% |

## 5.2 Nature of Crowd Capital Resources Being Generated in the Governance Context

As mentioned earlier in section # 2, the Crowd Capital resource can be generated in many forms, from IT-mediated Crowds, including knowledge, data, information, currency, ideas, creativity, task-work etc. Given the broad purview of the governance apparatus, we feel that it is useful to outline the specific sectors of governing within which the forms of Crowd Capital are being generated. Out of the 209 applications in our dataset, we find that IT-Mediated Crowds are being used to generate resources in a variety of governance areas. Table # 2 below summarizes this information for the top 10 most frequent governance contexts.

Similarly, while Figure # 4 summarizes the instances of Crowd Capital creation for all governance contexts in the dataset; in addition, we also explored the types of IT structures—episodic or collaborative—that facilitate the accumulation of Crowd Capital. The results are summarized below in Figure # 5. In particular, it is important to note the significant use of Episodic structures—sans social interaction in the crowd—for community improvement, environment, and Law Enforcement and the use of Collaborative IT, which requires interactions among the crowd participants, for generating legal Crowd Capital.

**Table # 2: Distribution of Crowd Capital Resource Types by Governance Sector**

| Crowd Capital Resource Generated by Governance Sector | % Of Distribution in Dataset |
|---|---|
| Community Improvement | 22% |
| Public Safety | 19% |
| Legal | 13% |
| Health Care | 12% |
| Transparency | 11% |
| Environment | 10% |
| Consultation | 6% |
| Agriculture | 3% |
| Military | 2% |
| Education | 2% |

**Figure # 4 – Crowd Capital Accumulated**

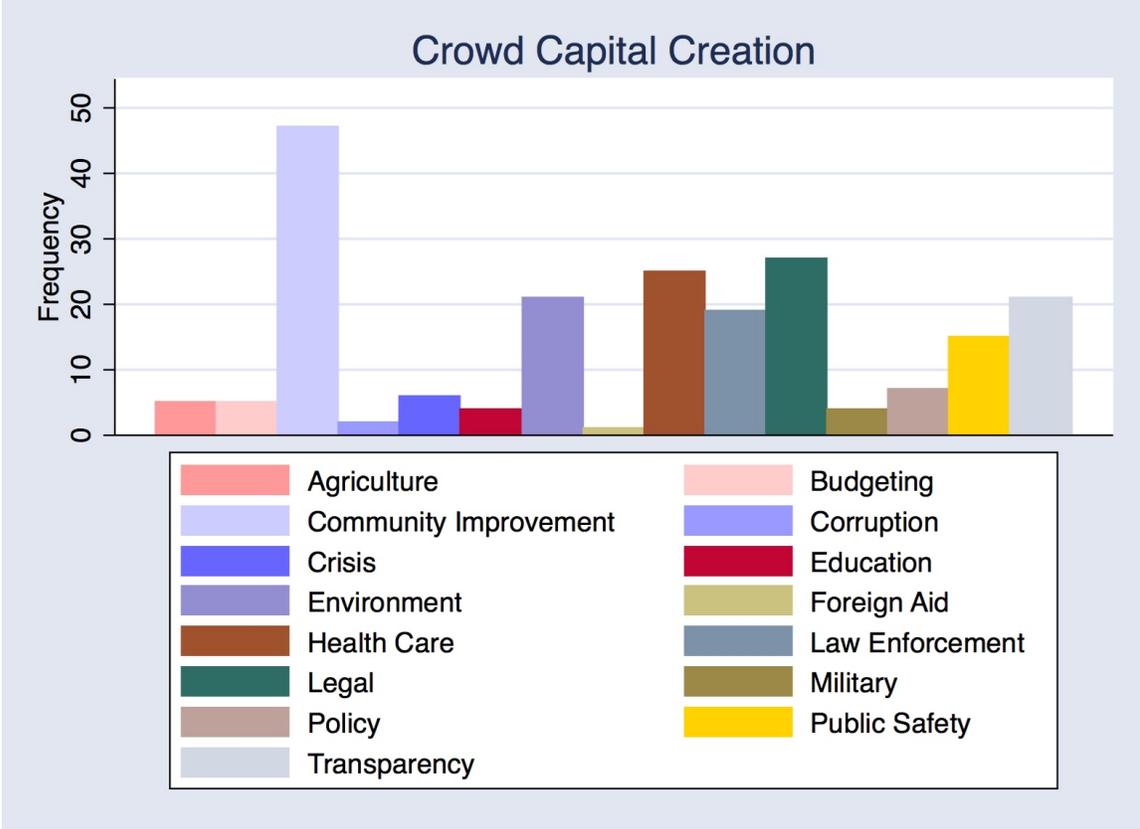

Figure # 5 – Percentage of Crowd Capital Accumulated using Episodic and Collaborative IT

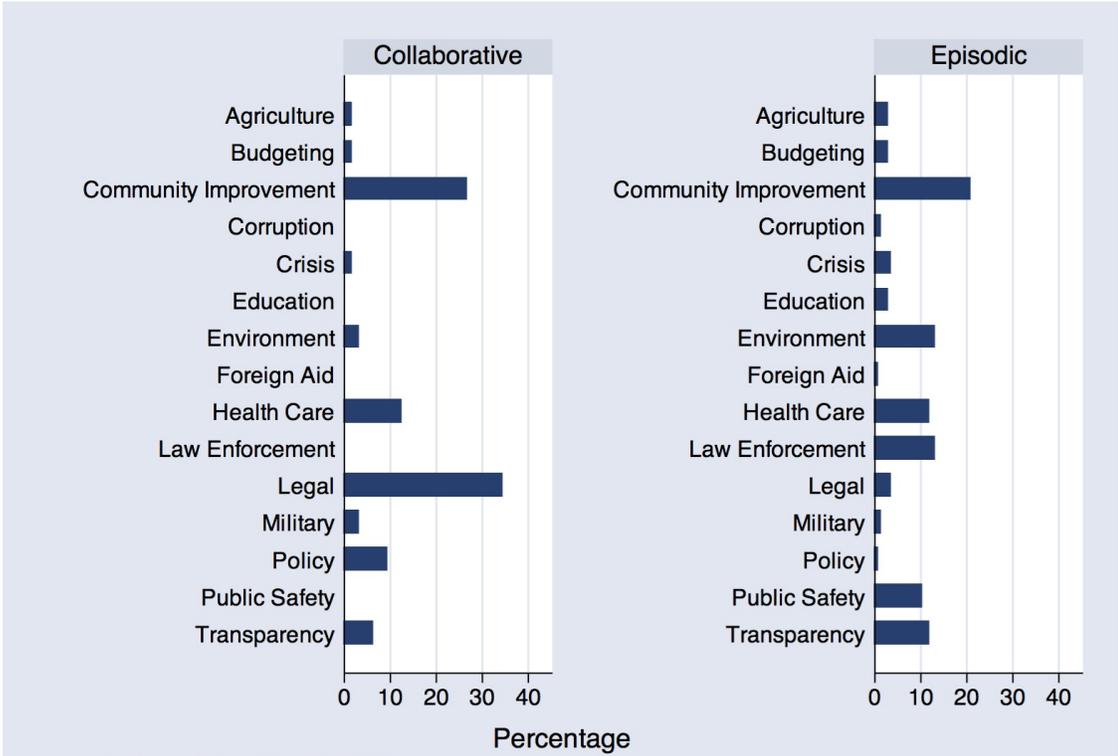

## 5.3 Level of Government Targeted by IT-Mediated Crowd Application

We assess the level of government target by the IT-mediated Crowd applications in our dataset. We distinguish between applications that solely target one level of government, for example municipal, state, and federal within a nation, multiple levels of within a nation, or transnational applications that target one or more levels of government in two or more nations. The results of this can be seen in Table # 3 below.

**Table # 3: Level of Government Target by Crowd Application**

| Level of Government Targeted by Application | Percentage of Distribution in Dataset |
|---|---|
| Transnational | 14% |
| National | 51% |
| Federal | 11% |
| State | 1% |
| Municipal | 23% |

## 5.4 IT-Structure of Applications in the Governance Context

As introduced in section # 2, the IT-structure of Crowd-IT is a crucial distinction. The choice of either episodic or collaborative IT-structures essentially determines the variety of dynamics the will exist between the implementing organization and the Crowd, and within the Crowd itself. Therefore, it is useful to understand the different IT-structures found to currently exist in the governance context. Of the 209 applications in our dataset, 69% were found to implement an episodic IT-structure, while 31% were found to engage Crowds through collaborative forms of IT-structure.

Furthermore, due to the large dataset availability, we are also able to gauge which type of technologies are more salient across different endeavors generating Crowd Capital. We find that while that while the web is used for generating all different types of Crowd Capital, mobile phones are salient in Law Enforcement and Community Development, whereas Software and SMS are most used in Health Care and Community Development respectively. These results are summarized in Figure # 6.

**Figure # 6 – Use of Crowd Capability in Crowd Capital Generation**

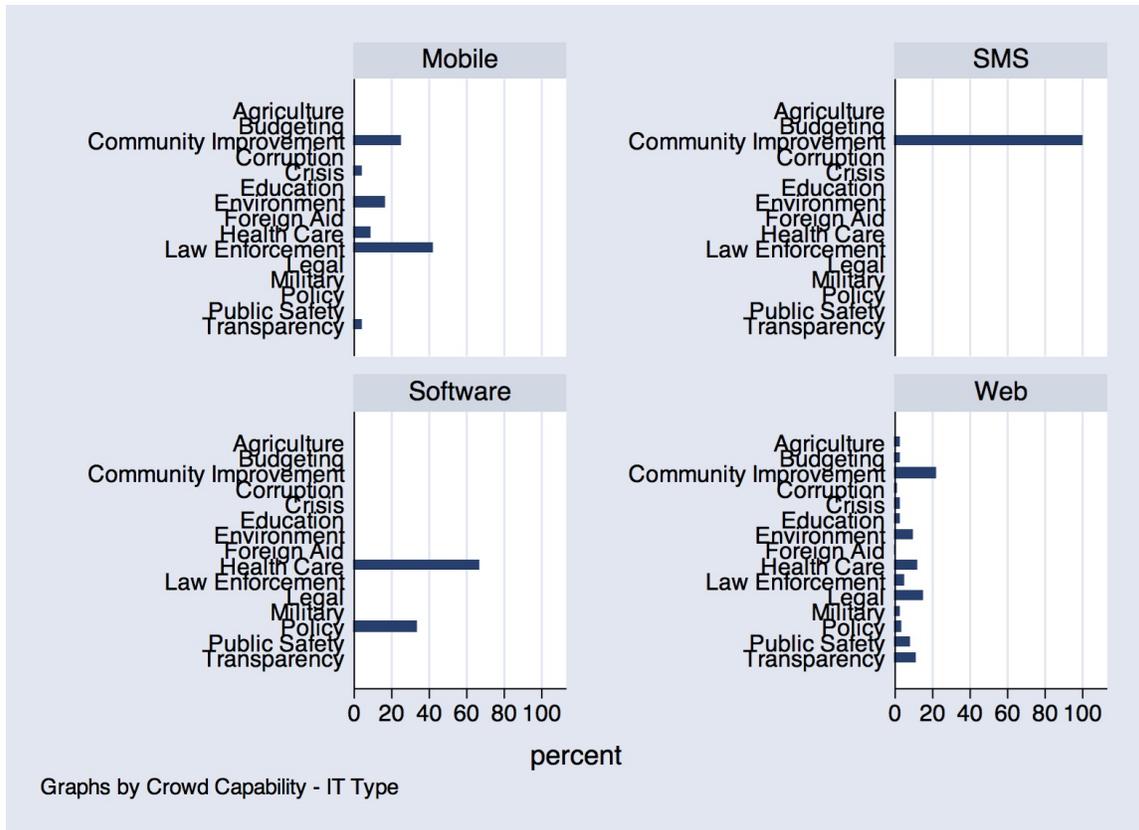

## 5.5 Geographic Location of Crowds Accessed in the Governance Context

We also assess the general regions in which the IT-mediated Crowd applications are currently functioning, to give us a sense of the global dispersion of the phenomenon, and the location of Crowds functioning in this respect.

We distinguish between applications that engage global crowds to generate governance resources from IT-mediated Crowds, as well as applications that target Crowds only in the following specific regions, detailed in Table # 4 below:

Table # 4:  Geographic Location of Crowds Accessed in Governance Context

| Geography of IT-Mediated Crowd Application Operations | % Of Distribution in Dataset |
|---|---|
| Global | 24% |
| Africa & Middle East | 8% |
| Europe & Russia | 13% |
| North America | 41% |
| South America | 1% |
| Asia | 10% |
| Oceania | 3% |

## 6. Discussion

The fundamental data collection and analysis that we have undertaken here raise some important and interesting questions on a number of fronts. In this section, we'll state and discuss these questions based upon our findings, and highlight some potential and observed implications of our analysis, in the hope of spurring future research and application of IT-mediated Crowd in governance contexts.

**6.1 Why does the use of IT-Mediated Crowds for Governance even exist?**

This question is not as spurious as it may initially seem. If nothing else, our work here illustrates that 209 projects have been started, and continue to operate on every continent around the world, and at every known level of government, to generate resources from IT-mediated Crowds for governance purposes. Federal agencies, Foreign services, Municipal governments, Transnational organizations, Non-profits, Social Enterprise organization and individuals, jointly and severally in numerous combinations, are acting to create and leverage Crowds for governance purposes.

Given that the private sector use of IT-mediated Crowds is where the phenomenon originated (see for example Crowdsourcing, Open Innovation platforms, and Crowdfunding) what does the recent transference of these ideas and potentials into the governance context, say about governance writ large? Is this a fad, or a sign of things to come? Are we in essence beginning to see a serious extension of the reach, expanse, importance, and influence of governance networks? Are these governance networks forming new socio-technical configurations of actors, issues, authority, legitimacy, and technologies?

Further, IT-mediated Crowds specifically engender new capabilities that represent a scale of individual participation, a speed and reach of knowledge creation, and massively parallel task work potentials that were previously not possible in our world, let alone readily available to most. We already live in a world, where issues routinely "go viral" (Zubiaga, Spina, Fresno, & Martínez, 2011), and in the process have already facilitated the toppling of numerous governments, such as in the Arab Spring (Lotan et al., 2011). Along similar lines, does the emergence of these applications signal the need for new consideration of the boundaries between public goods and private goods?

**6.2 All Governance Sectors All the Time?**

Our fundamental analysis in Table # 2 highlights the current distribution of Crowd Capital resources as being generated per governance sector. Therein, we highlight ten different sectors, more than half of which currently boast twenty or more applications in use around the globe. From public safety to the military, from the law to legislation, from health care and agriculture, to the environment, from public policy consultation to participatory budgeting applications, a litany of governance sectors are in essence being disrupted by numerous and disparate combinations of actors employing the potentials of IT-mediated Crowds, with little if any oversight. Are these sectors the low-hanging fruit? Or will this trend broaden and deepen?

**6.3 The New Civic Engagement?**

In Table # 2, the leading sector of the governance application of IT-mediated Crowds is what we term as Community Improvement endeavours, largely launched by municipalities, or municipally-focused actors to make real "rubber meets the road" improvement to local communities around the globe.

From fixing potholes, to adopting fire hydrants to the clean snow around them, reporting the incidence of graffiti, to providing ideas to make local communities better, are we seeing something of a new renaissance, or at least perhaps new forms of viable citizen engagement in civic affairs?

Further, ongoing initiatives like the Bloomberg Foundation's Mayoral Challenge, explicitly uses Crowdsourcing competitions to incentivize municipal-level leaders and bureaucrats to share their knowledge, experiences, and successes, with other cities. The net effect of such endeavours is to diffuse battle-tested ideas widely, in effect promoting a forum, and the resources, to share best of breed ideas far and wide. In much the same way that Smart City and Open Government Data initiatives have rapidly spread around the globe, are we at the beginning of a new wave of civic engagement through IT-mediated Crowds?

**6.4 Innovation as Governance**

An underlying, yet until now undiscussed implication of this work, is the fact that our dataset as a corpus essentially represents an in-depth study of IT innovation in the governance context. Innovation, for the most part considered a private-sector process, has now most certainly arrived in governance contexts, and has largely done so beyond the control of the government apparatus itself (with important exceptions like the US Federal government's continuing efforts, with initiatives like Challenge.gov).

Irrespective of how it has arrived, the idea that IT innovation should be an aim of governance, and that said IT innovations themselves should materially alter the dynamics, and processes of many governance sector themselves, seems to represent something of sea change in governance philosophy or possibility. Is this just a simple importation of private-sector values into government processes perceived as inefficient (surely we've heard that "small government" story before), or is something else going on?

When we consider that in the Innovation literature itself, that innovation is essentially a two-part process, first requiring invention, and then commercialization of said invention in a market, have we now reached a new paradigm of 'creative destruction' in the governance of societies? If so, what is being destroyed, and what is being created?

Clearly, as our work here illustrates, IT-mediated Crowd applications are being rapidly invented and commercialized/implemented in governance contexts, and this is occurring through sets of actors both endogenous to government (i.e. Politicians, Civil Services, Federal Agencies) and exogenous to government (Social Enterprises, Non-Profits, Foundations, Individuals etc).

It would seem that important questions remain unanswered in this domain.

## 7. Conclusion

In this work we have outlined a research program stemming from the compilation of the only-known dataset of endeavors implementing IT-mediated Crowds in the governance context. We illustrate some fundamental findings from our compiled dataset, illustrating numerous new and important findings in the process. From our investigation we learn the following basic facts about this salient new domain:

- There are at least 209 "pure play" Crowd Capital generating applications currently in use in the governance context, on every continent around the world, and at every level of government known to exist.

- In the governance context a wide variety of organizational actors are implementing IT-mediated Crowds, including Social Enterprises, Public/Private Partnerships, Politicians, Non-Profit organizations, and professional Civil Service organizations.

- More than $2/3^{rd}$ of Crowd Capital generating applications in the governance context use Episodic IT-structures to engage their IT-mediated Crowds.

- Crowd Capital resources are being generated from IT-mediated Crowds in at least ten different sectors of governance across the globe.

We extend these important fundamental contributions further by undertaking a discussion comparing these findings to our extensive knowledge of the literature on private-sector Crowd Capital generating endeavours, therein drawing-out observed and potentially important issues and implications of our data for researchers and practitioners alike.

Further, we contribute fundamentally to the literature on IT-mediated Crowds, by bringing this relatively large corpus of literature to bear on an important, new, growing, complex, and emerging context of governance, therein supplying the broadest and most holistic treatment that we are aware of merging IT-mediated Crowds and their use by and for governments and governance.

In a similar vein, we contribute fundamentally to the corpus of governance theory, by unpacking in detail new aspects of the network of actors in modern governance networks, detailing the new IT and analytical tools being used in said networks, and highlighting the sectors of governance where these applications predominate, both in terms of geographic location, and the sectors where these novel crowd-derived resources are being generated.

In sum, 1937, Coase posed some fundamental questions about organizations—one of them being "why do firms exist?"—and transactions cost theory was born. We have raised and strived to address similar fundamental questions in this work. We reason that we are the edge of a new wave of a governance paradigm that integrates IT-mediated crowds in its functioning and that the pervasiveness of such governance system where we engage with the crowd for a variety of governance needs will only increase. We have strived to showcase the emergence of the same through this exploratory work and we hope that the fundamental work undertaken here will assist both the research and governance practitioner communities in their effort to understand this salient shift in the governance context, and to improve the application of IT-mediated Crowds therein.